\documentstyle[pra,aps,epsf,multicol,rcs]{revtex}
\begin{document}
\RCS $Revision: 1.25.2.1 $
\RCS $Date: 1999/09/10 12:13:45 $
\draft
\title{Semiclassical analysis of level widths for one-dimensional potentials}
\author{Gert-Ludwig Ingold}
\address{Institut f\"ur Physik, Universit\"at Augsburg,
Universit{\"a}tsstra{\ss}e~1, D-86135 Augsburg, Germany}
\author{Rodolfo~A. Jalabert}
\address{Universit\'e Louis Pasteur, IPCMS-GEMME, 23 rue du Loess,
F-67037 Strasbourg Cedex, France}
\author{Klaus Richter}
\address{Max-Planck-Institut f{\"u}r Physik komplexer Systeme,
N{\"o}thnitzer Stra{\ss}e~38, D-01187 Dresden, Germany}
\date{September 10, 1999}
\maketitle
\widetext
\begin{abstract}
We present a semiclassical study of level widths for a class of one-dimensional 
potentials in the presence of an ohmic environment. Employing an expression 
for the dipole matrix element in terms of the Fourier transform of the 
classical path we obtain the level widths within the Golden rule 
approximation. It is found that for potentials with an asymptotic power-law 
behavior, which may in addition be limited by an infinite wall, the width that 
an eigenstate of the isolated system acquires due to the coupling to the
environment is proportional to its quantum number.
\end{abstract}


\raggedcolumns
\begin{multicols}{2}
\narrowtext
\section{Introduction}

Semiclassical approaches were essential at the advent of quantum mechanics
and have since remained a privileged tool for learning this 
subject, for developing our physical intuition on new problems and for 
performing analytical calculations \cite{Brack}. In particular, in the
one-dimensional (1d) case the semiclassical (WKB) approximation 
treats a wide class of confining potentials $V(q)$ in the limit of
large quantum numbers. We can write down closed expressions for eigenenergies
and eigenfunctions, and for power-law potentials the scaling of the
eigenenergies with the quantum numbers and the classical actions can be
established \cite{sukha73,carin93}. Once we have an almost complete solution of
a 1d problem we can use it as the basis for understanding more complicated
physical situations. This is the approach of this work, where we analyze the 
problem of a particle in $d=1$ coupled to a dissipative environment.

For an isolated quantum system, the ground state and the excited 
states are stable and the corresponding density of states is a 
series of delta functions. Any coupling to external degrees of freedom 
usually renders the excited states unstable, thus the levels acquire a 
finite lifetime \cite{weiss30} which is related to the level width through the 
time-energy uncertainty relation. The density of states of the 
system including the additional degrees of freedom still consists of (closely
spaced) delta functions, however
the reduced density of states corresponding to the 
system alone is smeared out compared to that of the uncoupled case. 

An example of such a situation is an atom coupled to the electromagnetic modes 
of the radiation field. If the atom were isolated from the field
there would be no transitions from excited states to states lower in energy.
This changes if we take the coupling to the radiation field into account. 
Then transitions between states may occur, and the finite lifetime of the 
excited states broadens the spectral lines associated with the transition. 

Two types of transitions are caused by the coupling to the radiation
field: stimulated processes and spontaneous emission. In the first case
photons have to be present, while in the second case the mere presence of 
field modes is sufficient to allow for transitions to energetically lower 
states accompanied by the emission of a photon. In the limit of zero 
temperature, the mean number of photons in the radiation field vanishes and 
the broadening of atomic levels is entirely due to spontaneous emission.

In the following, we replace the atom with a particle moving in a
one-dimensional potential. The environment is modeled by a set of
harmonic oscillators coupled bilinearly to the particle \cite{dittr98}. Using
semiclassical methods, we extend the scaling of 
Refs.~\cite{sukha73,carin93} to level widths and demonstrate that, for
a wide class of confining potentials, level widths are simply 
proportional to quantum number. To establish this general result 
we first present the theory for the level widths (Sec.~\ref{sec:gelw}) and
then apply it to a few specific examples:  harmonic oscillator 
(Sec~\ref{sec:osci}), 1d box (Sec.~\ref{sec:onb}),
half-oscillator (Sec.~\ref{sec:ho}) and Coulomb potential 
(Sec.~\ref{sec:rpcb}). The semiclassical approximation for level widths
(Sec.~\ref{sec:selw}) simplifies the calculations in the examples
considered and can be applied to general power-law potentials 
(Sec.~\ref{sec:stgpww}). In the concluding chapter (Sec.~\ref{sec:concl})
we analyze the experimental implications of our findings and their
possible extensions to higher dimensions.

\section{General expression for the level widths}
\label{sec:gelw}
As our model we consider a particle of mass $M$ moving in a one-dimensional
potential $V(q)$. The spectrum of the corresponding Hamiltonian
\begin{equation}
H_{\rm S}=\frac{p^2}{2M}+V(q)
\end{equation}
is assumed to consist of a discrete part at low energies which may be
followed by a continuous part at higher energies. It is on the discrete
part of the spectrum (consisting of eigenenergies $E_n$, $n=0,1,2,\dots$)
where we focus our analysis. To be consistent, we always denote 
the ground state by $n=0$ although this may lead to a slightly
unusual notation.

Since we consider the limit of large quantum numbers we require that 
the number of discrete eigenstates is infinite or at least can be made 
sufficiently large. This includes for example the radial part of the Coulomb 
problem but excludes the Morse potential. 

The eigenstates acquire a finite width if we weakly couple the particle 
to environmental degrees of freedom.  We assume that the environment
consists of a set of harmonic oscillators coupled bilinearly to the
particle. This leads to the full Hamiltonian
\begin{equation}
H = H_{\rm S} +
\sum_{j=0}^{\infty}\left[\frac{p_j^2}{2m_j}+\frac{m_j\omega_j^2}{2}
\left(x_j-\frac{c_j}{m_j\omega_j^2}q\right)^2\right]
\label{eq:htot}
\end{equation}
implying a coupling between system and environment through the
Hamiltonian
\begin{equation}
H_{\rm I} = -\sum_{j=0}^{\infty}c_jx_jq.
\end{equation}
By eliminating the environmental degrees of freedom, we obtain an effective
operator equation of motion \cite{dittr98,magal59},
\begin{equation}
\ddot q +\int_0^t{\rm d}s\gamma(t-s)\dot q(s)+\frac{1}{M}\frac{{\rm d}V}
{{\rm d}q} = \frac{1}{M}\xi(t) \ ,
\label{eq:reom}
\end{equation}
with damping kernel
\begin{equation}
\gamma(t) = \frac{2}{M}\int_0^{\infty}\frac{{\rm d}\omega}{\pi}
\frac{J(\omega)}{\omega}\cos(\omega t) \ ,
\label{eq:dampingkernel}
\end{equation}
spectral density of bath oscillators
\begin{equation}
J(\omega) = \pi\sum_{j=0}^{\infty} \frac{c_j^2}{2m_j\omega_j}\delta(\omega-\omega_j)
\ ,
\label{eq:density}
\end{equation}
and an operator-valued fluctuating force $\xi(t)$ which we do not need to 
specify further.

Of great importance is the special case of $J(\omega)=M\gamma\omega$. Then
the damping kernel becomes $\gamma(t)=2\gamma\delta(t)$. Noting that in  Eq.\
(\ref{eq:reom}) only half of the delta function contributes, the second term
becomes $\gamma\dot q(t)$ describing the  well-known classical
damping proportional to the particle velocity. This  type of damping
is often referred to as ohmic because such a damping term appears also in
equations describing electrical circuits containing an ohmic resistor.

The Hamiltonian (\ref{eq:htot}) provides a microscopic model for  
dissipation in quantum systems in the sense that dissipation is due to 
coupling to additional degrees of freedom. It does not, however, pretend that
in a real resistor we can identify  environmental oscillators
microscopically. 

Without going into  details, we mention that the Hamiltonian
(\ref{eq:htot}), besides the fact that the environment can be eliminated
analytically, provides a good description of many realistic systems. It has
been well studied over the years \cite{rubin60} and more recently became known
as Caldeira-Leggett model \cite{calde83} in the context of macroscopic quantum
phenomena.

Using the Hamiltonian (\ref{eq:htot}) and assuming weak coupling between
the particle and its environment, we calculate the zero temperature 
width of the $n$-th level by means of the Fermi golden rule
\begin{equation}
\Gamma_n = \frac{2\pi}{\hbar}\sum_{m,j=0}^{\infty}
\vert\langle m,1_j\vert H_{\rm I}\vert
n,0_j\rangle\vert^2\delta(E_n-E_m-\hbar\omega_j).
\end{equation}
This expression describes the decay of the state $n$ to an energetically
lower state $m$ by one excitation of the $j$-th environmental mode that
changes its occupation number from 0 to 1. 
Inserting the dipole matrix element
\begin{equation}
\langle 1_j\vert x_j\vert 0_j\rangle = \left(\frac{\hbar}{2m_j\omega_j}
\right)^{1/2}
\end{equation}
of the $j$-th environmental oscillator we get
\begin{equation}
\Gamma_n = \pi\sum_{m,j=0}^{\infty}
\frac{c_j^2}{m_j\omega_j}\vert d_{nm}\vert^2
\delta(E_n-E_m-\hbar\omega_j) \, 
\label{eq:gammamik}
\end{equation}
where
\begin{equation}
d_{nm} = \langle m\vert q\vert n\rangle
\end{equation}
is the dipole matrix element of the system.
The properties of the environmental modes appearing in Eq.\ (\ref{eq:gammamik})
can be expressed in terms of their spectral density (\ref{eq:density}), so we
write the level width as
\begin{equation}
\Gamma_n = \frac{2}{\hbar}\sum_{m=0}^{n-1}
\vert d_{nm}\vert^2J\left(\frac{E_n-E_m}{\hbar}
\right)\, .
\label{eq:gammagen}
\end{equation}
This result is valid for arbitrary bath density. Employing a cubic frequency 
dependence for $J(\omega)$ would lead to the natural decay width of an excited 
atomic state due to spontaneous emission, apart from prefactors arising from 
a proper treatment of the polarization of the emitted photons.

While the following calculations could in principle be performed for arbitrary
bath densities, we will consider only the important case of ohmic 
damping where the level widths are
\begin{equation}
\Gamma_n = \frac{2M\gamma}{\hbar^2}\sum_{m=0}^{n-1} 
\vert d_{nm}\vert^2(E_n-E_m) \, .
\label{eq:gammaohm}
\end{equation}
The sum over the system eigenstates is restricted since an
environment at zero temperature cannot excite the system into states
of higher energy. 

Eq.~(\ref{eq:gammaohm}) constitutes the starting point for a calculation of 
level widths that will be performed in the following sections. It represents, 
up to the factor $\gamma$, a sum over oscillator strengths 
\begin{equation}
f_{nm} = \frac{2M}{\hbar^2}\vert d_{nm}\vert^2(E_n-E_m).
\label{eq:oscistrength}
\end{equation}
The finiteness of the upper limit prevents us from
evaluating the level widths by the standard Thomas-Reiche-Kuhn
sum rule\cite{cohen78}
for oscillator strengths, 
\begin{equation}
\sum_{m=0}^\infty f_{nm} = 1.
\label{eq:trkrule}
\end{equation}

\section{Harmonic oscillator}
\label{sec:osci}
\begin{figure}
\begin{center}
\leavevmode
\epsfxsize=0.45\textwidth
\epsfbox{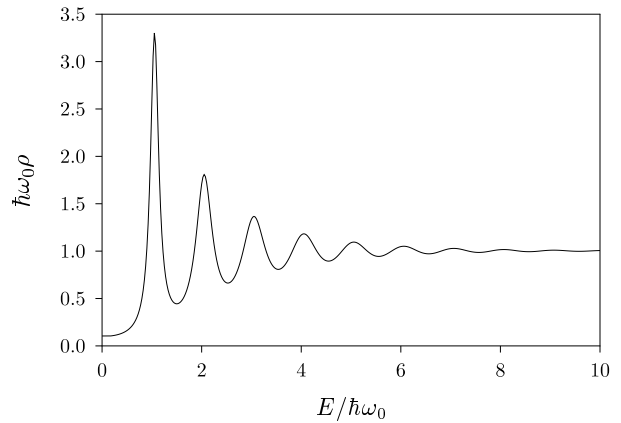}
\end{center}
\caption{Density of states for a harmonic oscillator of frequency $\omega_0$
coupled to an ohmic environment with $\gamma/2\omega_0=0.1$. The delta function
corresponding to the stable ground state is not shown.}
\label{fig:rho}
\end{figure}

The maybe simplest example is that of a harmonic potential with frequency 
$\omega_0$. The dipole matrix element in this case couples only nearest 
neighbors,
\begin{equation}
d_{nm} = \sqrt{\frac{\hbar}{2M\omega_0}}\left(\sqrt{n+1} \ \delta_{m,n+1}
+\sqrt{n} \ \delta_{m,n-1}\right).
\end{equation}
This leads to a simple and well-known result for the level widths of a 
damped harmonic oscillator \cite{cohen92}
\begin{equation}
\Gamma_n=n\gamma.
\label{eq:widthho}
\end{equation}
As an illustration we discuss the density of states for a damped
harmonic oscillator. In view of the relation between the partition function
of a canonical ensemble,
\begin{equation}
Z(\beta)=\int_0^{\infty}{\rm d}E\rho(E)\exp(-\beta E) \, ,
\label{eq:partdens}
\end{equation}
and the density of states $\rho(E)$, one may obtain the latter by inverse
Laplace transformation \cite{hanke95} from the partition function. 
For a damped harmonic oscillator, the partition function is\cite{weiss93}
\begin{equation}
Z(\beta) =\frac{1}{\hbar\beta\omega_0}\prod_{n=1}^{\infty}
\frac{\nu_n^2}{\nu_n^2+\nu_n\hat\gamma(\nu_n)+\omega_0^2}.
\label{eq:partfunc}
\end{equation}
Here, $\hat\gamma$ denotes the Laplace transform of the damping kernel
(\ref{eq:dampingkernel}) and $\nu_n=2\pi n/\hbar\beta$ are the Matsubara
frequencies. The product in 
Eq.~(\ref{eq:partfunc}) does not converge in the case
of an ohmic environment where $\hat\gamma(\nu_n)=\gamma$ and we therefore
introduce a cutoff in the spectral density of states. After inverse Laplace
transform, according to (\ref{eq:partdens}), we arrive at the density of 
states $\rho(E)$, for which the ohmic limit can be computed if the (infinite) 
ground state energy of system plus environment is subtracted. The result is 
shown in Fig.~\ref{fig:rho} for ohmic damping with $\gamma/2\omega_0=0.1$,
omitting the delta function corresponding to the stable ground state. The 
levels broaden with increasing energy and for large energies only the average 
density of states $\rho=1/\hbar\omega_0$ is seen.

According to Eq.~(\ref{eq:widthho}) the level widths for the harmonic 
oscillator are proportional to $n$ or, equivalently, to $E_n$. 
In general, however, $E_n$ is not proportional to $n$. Then the question
arises whether the widths are simply determined by $n$, or
by $E_n$ and whether or not there is a universal behavior.

\section{One-dimensional box}
\label{sec:onb}
A more non-trivial example is that of a particle in a one-dimensional box 
of length $L$. The eigenenergies are 
\begin{equation}
E_n = \frac{\hbar^2\pi^2}{2ML^2}(n+1)^2
\end{equation}
and dipole matrix elements can be easily calculated.
For reasons of symmetry a dipole coupling can induce
transitions only between states of different parity. For $n-m$ odd one
obtains
\begin{equation}
d_{nm} = -\frac{8L}{\pi^2} \ \frac{(n+1)(m+1)}{[(n+1)^2-(m+1)^2]^2} \ .
\end{equation}
We are interested in the universal behavior 
for large quantum numbers, where $d_{nm}$ decreases as $(n-m)^2$. Therefore,
we define $l=n-m$ and write approximately for $l\ll m,n$
\begin{equation}
d_{nm} = -\frac{2L}{\pi^2}\frac{1}{l^2} \ .
\label{eq:diboxln}
\end{equation}

\begin{figure}
\begin{center}
\leavevmode
\epsfxsize=0.45\textwidth
\epsfbox{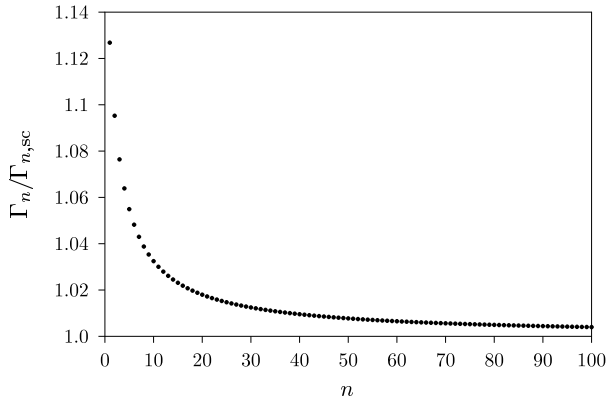}
\end{center}
\caption{Ratio between the result (\ref{eq:boxres}) and the exact width
$\Gamma_n$ for a particle in a box as a function of the state $n$.}
\label{fig:boxwn}
\end{figure}

We can now evaluate the level widths for large quantum
number $n$ by assuming that the sum over all final states $m$ converges fast
enough that only differences $l \ll  n,m$ are important. Working to
leading order in $l/n$ we extend the sum over $l$ to infinity and
obtain, with Eq.~(\ref{eq:gammaohm}),
\begin{equation}
\Gamma_{n, {\rm sc}} = \frac{8\gamma}{\pi^2} \ n \ \sum_{k=0}^{\infty}
\frac{1}{(2k+1)^3}.
\end{equation}
The sum can be expressed in terms of Riemann's zeta function as
$(7/8)\zeta(3) = 1.051\dots$ \cite{abram72}. The level widths at large 
quantum numbers are therefore given by
\begin{equation}
\Gamma_{n, {\rm sc}}=0.852\dots \gamma n.
\label{eq:boxres}
\end{equation}
The subindex ``${\rm sc}$'' stands for ``semiclassical'' since, as  seen 
in the next section, Eq.~(\ref{eq:boxres}) results directly  from a WKB
approximation. The quality of our approximate result is seen in
Fig.~\ref{fig:boxwn}
where we show the ratio between the exact width $\Gamma_n$ and the
semiclassical width $\Gamma_{n, {\rm sc}}$. The relative error drops below
1\% at about $n=38$.

The result (\ref{eq:boxres}) indicates that the proportionality in the case 
of the harmonic oscillator should be interpreted to be to the quantum number 
and not to the eigenenergy. The prefactor resulting for the 1d box is different 
from that of the harmonic oscillator. Therefore a potentially universal 
behavior can only concern the scaling with the quantum number, while the 
prefactor will be specific to the given potential.

\section{Semiclassical expression for the dipole matrix element}
\label{sec:selw}
In the two examples treated above, the harmonic oscillator and the
1d box, the wave functions are known and the dipole matrix elements
are readily calculated. To treat  generic potentials in the
limit of large quantum numbers we  use a semiclassical
approximation for the dipole matrix element. Such an expression
(see Eq.~(\ref{eq:dipolsemi}) below) was first derived by Debye
\cite{debye27} and we  present it 
in this section to emphasize the physical assumptions on which 
it is based. 

For a 1d confining potential $V(q)$ the WKB approximation \cite{gutz90} for 
the wave function  with eigenenergy $E_n$ ($n \gg 1$) is
\begin{equation}
\psi_{n, {\rm sc}}(q)
=\left(\displaystyle\frac{4}{T}\frac{M}{p(q,E_n)}\right)^{1/2} 
\cos\left(\displaystyle\frac{1}{\hbar}S(q,E_n)-
\frac{\pi}{4}\right)\, .
\label{eq:swf}
\end{equation}
Here, $q$ lies between the two turning points $q_1$ and $q_2$
($V(q_1)=V(q_2)=E_n$, $q_1<q_2$) which determine the limits of the
classical motion at the energy $E_n$. The classical action is
computed from $q_1$ to $q$ according to
\begin{equation}
S(q,E_n) = \int_{q_1}^{q} {\rm d}q\, p(q,E_n)
\label{eq:ca}
\end{equation}
with
\begin{equation}
p(q,E_n) = \left[2M(E_n-V(q))\right]^{1/2}
\end{equation}
The quantization condition determining $E_n$ depends on the action 
along a period of the classical motion
\begin{equation}
S(E_n) = 2 S(q_2,E_n) = 2\pi \hbar\left(n+\frac{1}{2}\right) \, ,
\label{eq:qc}
\end{equation}
and the period $T$ is given by
\begin{equation}
T(E) = \frac{{\rm d}S}{{\rm d}E} = \int_{q_1}^{q_2} {\rm d}q
\frac{2M}{p(q,E_n)} \ .
\label{eq:pcm}
\end{equation}

The commutation relation $({\rm i}/\hbar)[H,q]=p/M$ allows us to express the 
dipole matrix element in terms of the momentum matrix element.
We can then write
\begin{equation}
d_{nm} = \frac{\hbar^2}{M(E_n-E_m)} \ \int_{-\infty}^{+\infty} {\rm d}q \
\psi_{m, {\rm sc}}^{*}(q) \ \frac{\partial}{\partial q} 
\psi_{n, {\rm sc}}(q) \ .
\label{eq:dnm1}
\end{equation}
Under the assumption $l=n-m \ll n,m$, both eigenenergies have almost the same 
classically allowed region. Neglecting the contribution of the classically 
forbidden region, we restrict the integration to the interval $(q_1,q_2)$. 
Moreover, we assume that $V(q)$ is a smoothly varying function, and thus 
the derivative with respect to $q$ is carried out only in the cosine, which has
a small factor $\hbar$ in its denominator, but not in the prefactor. With these
assumptions we have
\begin{eqnarray}
\label{eq:dnm2}
d_{nm} &=& -\frac{2\hbar}{T(E_n-E_m)} \ \int_{q_1}^{q_2} {\rm d}q  \nonumber \\
&&\times\left\{\sin{\left[\frac{1}{\hbar}\left(S(q,E_n)-S(q,E_m)\right)
\right]} \right.  \\
&&\hphantom{\times(}-\left. \cos{\left[\frac{1}{\hbar}
\left(S(q,E_n)+S(q,E_m)\right) \right]} \right\} \ . \nonumber
\end{eqnarray}
The second term in the integral is rapidly oscillating and is
negligible upon integration. For the first term we use  the relation
\begin{equation}
S(q,E_n)-S(q,E_m) \simeq t(q,E_n)(E_n-E_m) \ ,
\label{eq:diffact}
\end{equation}
where $t(q,E_n)$ is the time for a classical particle
to move from $q_1$ to $q$ at energy $E_n$. With Eqs.~(\ref{eq:qc})
and (\ref{eq:pcm}) we can write
\begin{equation}
E_n-E_m \simeq \frac{\partial E}{\partial n} l = \frac{2\pi \hbar}{T} l\ ,
\label{eq:diffener}
\end{equation}
and then find from Eq.~(\ref{eq:dnm2})
\begin{eqnarray}
\label{eq:dipolsemi}
d_{n,n-l} &=&  - \frac{1}{\pi l} \int_{q_1}^{q_2} {\rm d} q  
\sin{\left(2\pi l \frac{t(q)}{T}\right)} \nonumber \\
&=& \frac{1}{T(E)} \int_{0}^{T(E)} {\rm d}t q(t,E) 
\cos{\left(2\pi l \frac{t}{T(E)}\right)}
\end{eqnarray}
where the second line has been obtained by means of the substitution 
${\rm d}q = {\dot q}{\rm d}t$ and partial integration. 

Eq.\ (\ref{eq:dipolsemi}) constitutes the interesting result found by Debye 
which allows to relate the dipole matrix elements in the semiclassical limit 
to the Fourier components of the classical motion of the particle. It extends
Bohr's correspondence principle which states that in the limit $\hbar\to 0$ the
frequencies of an atomic transition should agree with electrodynamics. By
virtue of Eq.\ (\ref{eq:dipolsemi}), a correspondence between quantum theory and
electrodynamics can not only be established for frequencies but also for 
intensities which according to Eq.\ (\ref{eq:oscistrength}) are related to the 
square of dipole matrix elements.

Eq.~(\ref{eq:dipolsemi}) is the leading order approximation in 
$\hbar$. Higher order corrections have been derived in Ref.\
\cite{karrl98}; however we will make no use of them in this work.

In the presence of hard walls the WKB approximation is still
applicable and only a change in the phase of the semiclassical
wavefunction (\ref{eq:swf}) is needed to recover the result for the 
1d box. Then the semiclassical approximation to the dipole matrix
element is  the same as for smooth potentials (Eq.~(\ref{eq:dipolsemi})).
The period of a classical trajectory with energy $E$ in the box is 
\begin{equation}
T(E) = L\left(\frac{2M}{E}\right)^{1/2} \ ,
\end{equation}
leading in the limit of large $n$ to the forms (\ref{eq:diboxln}) and
(\ref{eq:boxres}) for the matrix elements and the level widths, respectively.

\section{Half-oscillator}
\label{sec:ho}
Our next example is the harmonic oscillator with an infinite wall at
$q=0$. The particle is therefore confined to $q\ge0$. In contrast to the usual 
harmonic oscillator, the dipole matrix element no longer couples only nearest 
neighbor states. The eigenstates of the half-oscillator are given by the odd
eigenstates of the harmonic oscillator with the prefactor adjusted to
account for the restricted interval of normalization. The dipole matrix
element $d_{nm}$ may then be evaluated by expressing the Hermite polynomial 
with the higher quantum number $n > m$ by means of the Rodrigues formula,
\begin{eqnarray}
d_{nm} &=& -\left(\frac{\hbar}{M\omega\pi}\right)^{1/2} 
\frac{1}{2^{n+m}[(2n+1)!(2m+1)!]^{1/2}}\nonumber\\
&&\times\int_0^{\infty}{\rm d}\xi
\xi H_{2m+1}(\xi)\frac{{\rm d}^{2n+1}}{{\rm d}\xi^{2n+1}}\exp(-\xi^2)\ .
\end{eqnarray}
After repeated partial integration and use of
\begin{equation}
\left.\frac{{\rm d}^k}{{\rm d}\xi^k}\left(\xi H_{2m+1}(\xi)\right)\right
\vert_{\xi=0} = (-1)^{m+1-k/2}\frac{2^{k-1}k(2m+1)!}{\displaystyle
\left(m+1-k/2\right)!}\ ,
\end{equation}
which holds for $k$ even and yields zero for $k$ odd, one arrives at
\begin{eqnarray}
d_{nm} &=& 2\left(\frac{\hbar}{M\pi\omega}\right)^{1/2}\left(
\frac{(2m+1)!}{2^{2(m+n+1)}(2n+1)!}\right)^{1/2}\nonumber\\
&&\times(-1)^{n-m+1}
\sum_{l=1}^{m+1}2^{2l}l\frac{[2(n-l)]!}{(m-l+1)!(n-l)!}\ .
\end{eqnarray}
The sum may be evaluated to obtain
\begin{eqnarray}
d_{nm} &=& \left(\frac{\hbar}{M\pi\omega}\right)^{1/2}\frac{(-1)^{n-m+1}}
{2^{n+m-1}}\frac{\left[(2n+1)!(2m+1)!\right]^{1/2}}{n!m!}\nonumber\\
&&\times\frac{1}{4(n-m)^2-1}.
\end{eqnarray}

A corresponding semiclassical evaluation of the dipole matrix element using
Eq.~(\ref{eq:dipolsemi}) leads to 
\begin{equation}
d_{n,n-l} = -\frac{4}{\pi}\left[\frac{\hbar}{M\omega}
\left(n-\frac{1}{4}\right)\right]^{1/2}\frac{1}{4l^2-1}\ .
\end{equation}
For $n\gg l$ this agrees with the exact result up to an irrelevant sign. For 
the level widths we then obtain
\begin{equation}
\Gamma_{n, {\rm sc}} = \frac{8}{\pi^2}\gamma n = 0.810\dots\gamma n
\end{equation}
and thus again a proportionality to the quantum number.
Even though the eigenenergies of the one-dimensional box and the
half-oscillator depend differently on the quantum number and in spite of
the fact that the dipole matrix elements of oscillator and
half-oscillator are quite different, the final result for the level
widths is proportional to the quantum number $n$ in both cases. 

\section{Radial part of Coulomb potential}
\label{sec:rpcb}
A less trivial dependence of the dipole matrix elements on quantum 
number is found for our last example,  the radial part of 
the three-dimensional Coulomb problem for vanishing angular momentum. In this 
case, an infinite potential wall again restricts the particle to positive 
$q$ where it  moves in an attractive Coulomb potential $-A/q$. Unlike
the previous examples, the spectrum of the Coulomb problem contains a 
continuous part at positive energies. The discrete part of the spectrum 
consists of the eigenenergies
\begin{equation}
E_n=-\frac{MA^2}{2\hbar^2}\frac{1}{(n+1)^2}.
\label{eq:cenergy}
\end{equation}

Dipole matrix elements can be obtained by solving the
Schr{\"o}dinger equation in momentum representation \cite{ivash72} and read
\cite{sussk88}
\begin{equation}
d_{nm} = \frac{8\hbar^2}{M\pi A}\frac{[(n+1)(m+1)]^{5/2}}{(n+1)^2-(m+1)^2} 
I_{nm}
\end{equation}
with
\begin{equation}
I_{nm} = \int_0^{\infty}{\rm d}u \frac{u\sin[f_n(u)-f_m(u)]}
{[1+(n+1)^2u^2][1+(m+1)^2u^2]}
\end{equation}
and
\begin{equation}
f_n(u) = 2(n+1)\mathrm{arctan}[(n+1)u].
\end{equation}
For $n,m\gg 1$ and $n>m$ this becomes \cite{gores82}
\begin{equation}
d_{nm} = (-1)^{n-m+1} \frac{2^{4/3}3^{1/6}\hbar^2}{M\pi
A}\Gamma\left(\frac{2}{3}\right)\frac{(nm)^{11/6}}{(n^2-m^2)^{5/3}}.
\label{eq:cdipol}
\end{equation}

It is instructive to derive this result from the semiclassical
expression (\ref{eq:dipolsemi}). Even though the trajectory $q(t)$ cannot be
obtained explicitly, we can obtain the matrix elements by
observing that the behavior of the Fourier coefficients of $q(t)$ 
is dominated by the most singular part of the original function which
in our case is the motion around the reflection point at $q=0$.

Energy conservation leads to the trajectory at energy $E$ in 
implicit form as
\begin{equation}
\left(\frac{2}{M}\right)^{1/2}\frac{\vert E\vert^{3/2}}{A}t +
\frac{\pi}{2} = {\rm arcsin}(q')-q'^{1/2}(1-q')^{1/2}
\label{eq:couleom}
\end{equation}
with $q' = (\vert E\vert/A)q$. From this result we compute the period 
\begin{equation}
T = \pi\left(\frac{M}{2}\right)^{1/2}\frac{A}{\vert E\vert^{3/2}}
\end{equation}
and by expanding the arcsin we find, for the trajectory around the reflection
point encountered at $t=0$,
\begin{equation}
q = \left(\frac{9A}{2M}\right)^{1/3}t^{2/3}.
\end{equation}
We then write the trajectory for $0\le t<T$ as
\begin{equation}
q = \left(\frac{9A}{2M}\right)^{1/3}t^{2/3}\left(1-\frac{t}{T}\right)^{2/3}
+q_{\rm reg}
\label{eq:ctraj}
\end{equation}
where the first term reproduces the most singular behavior of the
trajectory around $t=0$ and $t=T$ while the second term contains the
more regular part and is of no further interest. From Eq.~(\ref{eq:ctraj})
we find with Eq.~(\ref{eq:dipolsemi}) for $n\gg l\gg1$ and up to an irrelevant
sign
\begin{equation}
d_{n,n-l} = \frac{\hbar^2}{M\pi A}\frac{3^{1/6}}{2^{1/3}}\Gamma\left(
\frac{2}{3}\right)\frac{n^2}{l^{5/3}}.
\label{eq:cdipols}
\end{equation}
Written in symmetric form this agrees with Eq.~(\ref{eq:cdipol}) in the limit
$n\gg l$.

The level widths can now be obtained from Eq.~(\ref{eq:gammaohm}) with 
(\ref{eq:cenergy}) and (\ref{eq:cdipols}) as
\begin{equation}
\Gamma_{n, {\rm sc}} = \frac{6^{1/3}[\Gamma(2/3)]^2}{\pi^2}
\zeta\left(\frac{7}{3}\right) \gamma n = 0.477\dots \gamma n \ .
\end{equation}
Again, we find the proportionality to the quantum number $n$ which, in view
of the complicated structure of the dipole matrix element, may come as a
surprise. This indicates that this proportionality might be more universal.
We will indeed prove it in the following section for power-law potentials
with and without walls. 

\section{Semiclassical treatment for power-law potentials}
\label{sec:stgpww}
In this section we generalize the previous results to power-law potentials 
of the form $V(q) = A|q|^{\alpha}$. The amplitude $A$ and the exponent $\alpha$
should have opposite signs. The possible coordinates might be limited by an
infinite potential wall, and we assume that such a wall is present at $q=0$
whenever $\alpha<0$. The case $\alpha=0$ will be excluded because it requires
two walls and thus becomes the box potential already treated in 
Sec.~\ref{sec:onb}. Furthermore, we have to restrict the exponent to 
$\alpha>-2$. At $\alpha=-2$ the action becomes independent of energy as will 
become clear from Eq.\ (\ref{eq:scal_qc}) below so that we have to exclude this 
pathological case. We emphasize, that an attractive $1/q^2$ potential can never 
appear in an radial equation of motion in $d>1$ where the angular degrees of 
freedom have been eliminated.

Since our semiclassical approach requires sufficiently large energies,
the discussion applies also to potentials which effectively behave like a 
power-law potential at higher energies, e.g.\ the quartic double-well 
potential.

According to Eq.~(\ref{eq:gammaohm}) the $n$ dependence of the level width 
is determined from the 
dipole matrix element $d_{nm}$ and the energy difference $(E_n-E_m)$. 
Within the hypothesis established in Sec.~\ref{sec:selw} (which was more
general than for power-law potentials) we saw that  $(E_n-E_m)$ is simply
inverse proportional to the period of the classical motion between the
turning points (Eq.~(\ref{eq:diffener})). 

To evaluate the dependence on $n$ of the level widths
(\ref{eq:gammaohm}) we
employ scaling properties of the energy conservation condition
\begin{equation}
E=\frac{M}{2}\dot q^2 + A q^{\alpha} \, .
\label{eq:encon}
\end{equation}
Introducing a dimensionless coordinate 
\begin{equation}
q' = \displaystyle \left(\frac{\vert A\vert}{E}\right)^{1/\alpha}q 
\end{equation}
as in (\ref{eq:couleom}) and dimensionless time
\begin{equation}
t' = \frac{A^{1/\alpha}E^{(\alpha-2)/2\alpha}}{M^{1/2}}t \ ,
\label{eq:time_scal}
\end{equation}
(\ref{eq:encon}) becomes
\begin{equation}
1=\frac{1}{2}{\dot q}'^2 + {\rm sign}(A)\ {q'}^{\alpha} 
\label{eq:scal_hamilton}
\end{equation}
where ${\dot q}' = {\rm d}q'/{\rm d}t'$ and ${\rm sign}(A)$ denotes the sign 
of $A$. The quantization condition (\ref{eq:qc}) reads in scaled variables 
\begin{equation}
S(E) = \frac{M^{1/2}E^{(2+\alpha)/2\alpha}}{\vert A\vert^{1/\alpha}}
\oint {\rm d}q'\ {\dot q}'
 =  2\pi \hbar n\ ,
\label{eq:scal_qc}
\end{equation}
where the integral runs over one period. On the right-hand-side we have omitted
an $n$-independent term which depends on the presence of walls in the
potential.
With $S' = \oint {\rm d}q' \dot q'$ we then find for the energy eigenvalues
\begin{equation}
E = \left[\left(\frac{2\pi\hbar}{S'}\right)^{2\alpha}\frac{\vert A\vert^2}
{M^{\alpha}}\right]^{1/(2+\alpha)} n^{2\alpha/(2+\alpha)} \ .
\label{eq:en}
\end{equation}
This is in agreement with previous results for power-law potentials 
\cite{sukha73,carin93} where $S'$ has been evaluated explicitly. On the other
hand, Eq.\ (\ref{eq:en}) is still correct for sufficiently large $n$ if the
potential behaves only asymptotically like a power-law. Then, in general,
$S'$ can no longer be evaluated analytically.

The semiclassical dipole matrix element (\ref{eq:dipolsemi}) reads
after scaling 
\begin{equation}
d_{n,n-l} =  
\displaystyle \left(\frac{E}{\vert A\vert}\right)^{1/\alpha}d_l'  
\label{eq:scale_dipole}
\end{equation}
with
\begin{equation}
d_l' =  
\displaystyle -\frac{1}{\pi l} 
\int_{q'_1}^{q'_2}{\rm d}q'\ \sin\left(2\pi l \frac{t'(q')}{T'}\right)
\label{eq:dipol_scal}
\end{equation}
where $T'$ denotes the period $T$ scaled according to (\ref{eq:time_scal}).

In view of Eqs.~(\ref{eq:pcm}), 
(\ref{eq:en}) and (\ref{eq:scale_dipole}), the level width 
(\ref{eq:gammaohm}) can be expressed as
\begin{equation}
\Gamma_{n, {\rm sc}} = \gamma\frac{8\pi^2}{S'^2}\frac{2+\alpha}{2\alpha} 
\left[\sum_{l=1}^n l{d'_l}^2\right]n \ .
\label{eq:central_fast}
\end{equation}
Apart from the factor $n$ this result still depends on the state number $n$ via
the upper limit of the sum. As a last step, we therefore have to consider the
convergence properties of this sum.

For $\alpha>0$ and in the absence of an infinite potential wall, the particle
velocity is always continuous and, as a consequence, the scaled dipol moment 
(\ref{eq:dipol_scal}) will decay at least as $l^{-2}$. This still holds for
a wall with finite potential to its one side because in the worst case the
reflection will lead to a triangular cusp singularity in the trajectory. In 
this case, the dipol moment will decay as $l^{-2}$. 

The case of negative exponent $\alpha$ with an infinite potential wall at the
divergence at $q=0$ is more interesting. Close to $q=0$ we may neglect the
constant on the left-hand-side of (\ref{eq:scal_hamilton}). Assuming that the
reflection happens at $t'=0$ we find for the trajectory close to the reflection
point $q'\sim \vert t'\vert^{2/(2-\alpha)}$. Proceeding as in Sec.\ 
\ref{sec:rpcb}, one finds for the scaled dipole matrix element
$d_l'\sim l^{-(4-\alpha)/(2-\alpha)}$. For $\alpha=-1$ this is in agreement 
with (\ref{eq:cdipols}). For $\alpha>-2$ it follows that $d_l'$ decays always
faster than $l^{3/2}$.

As a consequence, the argument of the sum in Eq.\ (\ref{eq:central_fast}) decays 
faster than $1/l^2$ for all potentials under consideration. Neglecting terms of
order $1/n$, as is consistent with our previous approximations, we may extend
the upper summation limit to infinity and arrive at our final result
\begin{equation}
\Gamma_{n, {\rm sc}} = \gamma\frac{8\pi^2}{S'^2}\frac{2+\alpha}{2\alpha} 
\left[\sum_{l=1}^{\infty} l{d'_l}^2\right]n \ .
\label{eq:central}
\end{equation}
For sufficiently large energies, the level width is therefore proportional to 
the state number $n$. We point out that the proportionality constant depends on 
$\alpha$ and $\gamma$ but not on $M$ and $A$. While these properties are special
to the case of ohmic damping, an extension to other bath densities would be
straightforward along the lines presented here.

\section{Conclusions}
\label{sec:concl}

Semiclassical methods for radiative lifetimes, ubiquitous in spectroscopic 
measurements, have mainly been derived for excited states of hydrogenlike 
atoms\cite{atoms}. Here, we applied such methods to a whole class of
one-dimensional potentials which behave like a power-law for large energies
and may contain one or two walls. Using a semiclassical expression for the 
dipole matrix element in terms of the Fourier transform of the classical paths 
we have shown that for ohmic damping the level width is proportional to the 
number $n$ of the state. This result is valid for sufficiently large $n$, 
therefore requiring weak coupling to the environment in order to be observable.
The proportionality of level widths with the state number of the harmonic 
oscillator could thereby be generalized to a large class of one-dimensional 
potentials. The prefactor of the linear law depends on the specific potential.

Extensions of our results to higher dimensions are interesting since
we will be approaching experimental reality and  will be able to 
investigate the relevance of the integrability of the classical system on
level widths. The chaotic or integrable character of the classical dynamics
has been shown to be of crucial importance for other observables like the
orbital magnetic response \cite{physrep}. 

The problem of an integrable 2d system, where we can separate the dynamics,
can be reduced to two 1d problems. For the 2d box it is readily  
demonstrated that the {\em mean} behavior of $\Gamma_n$ is again
a linear increase with $n$. 

The problem becomes more involved in chaotic systems
where the energy is the only conserved quantity and the explicit forms of
individual wavefunctions are not known. When increasing the number of degrees 
of freedom of the system from 1 to 2 we start seeing fluctuations of
level widths with respect to a secular behavior. Since level widths are 
 given by matrix elements of the dipole operator we expect
that the fluctuations will be more important in the integrable case, 
where selection rules  impose wide variations according to the 
quantum numbers of the initial and final states. 

\acknowledgements
We have benefited from discussions with J.-Y.\ Bigot, H.\ Grabert, S.\ Kohler, 
and S.\ Otto. Part of this work has been carried out while one of us (GLI) was 
at the Centre d'Etudes de Saclay with financial support from the 
Volkswagen-Stiftung.

\end{multicols}
\end{document}